\documentclass[twocolumn,aps,showpacs,preprintnumbers,amsmath,amssymb]{revtex4}

\usepackage[dvips]{graphicx}
\usepackage{epsfig}
\usepackage{amsmath}
\usepackage{subfigure}
\usepackage{color}
\usepackage{ulem}
\usepackage{float}
\usepackage{color}

\begin{document}

\title{Performance analysis of an interacting quantum dot thermoelectric set up}
\bigskip

\author{Bhaskaran Muralidharan and Milena Grifoni}
\affiliation{Institut f\"ur Theoretische Physik, Universit\"at Regensburg, Regensburg D-93040, Germany}
\date{\today}
\medskip

\widetext
\begin{abstract}
In the absence of phonon contribution, a weakly coupled single orbital non-interacting quantum dot thermoelectric set up is known to operate reversibly as a Carnot engine. 
This reversible operation, however, occurs only in the ideal case of vanishing coupling to the contacts, wherein the transmission function is delta-shaped, and 
under open-circuit conditions, where no electrical power is extracted. In this paper, we delve into the thermoelectric performance of quantum dot systems by analyzing the power output and efficiency 
directly evaluated from the non-equilibrium electric and energy currents across them. In the case of interacting quantum dots, 
the non-equilibrium currents in the limit of weak coupling to the contacts are evaluated using the Pauli master equation approach. 
The following fundamental aspects of the thermoelectric operation of a quantum dot set up are discussed in detail: 
a) With a finite coupling to the contacts, a thermoelectric set up always operates irreversibly under open-circuit conditions, with a zero efficiency.
b) Operation at a peak efficiency close to the Carnot value is possible under a finite power operation. In the non-interacting single orbital case, 
the peak efficiency approaches the Carnot value as the coupling to the contacts becomes smaller. In the interacting case, this trend depends non-trivially on the interaction parameter $U$. 
c) The evaluated trends of the maximum efficiency derived from the non-equilibrium currents deviate considerably from the conventional {\it{figure of merit}} $zT$ based results. 
Finally, we also analyze the interacting quantum dot set up for thermoelectric operation at maximum power output.
\end{abstract}
\maketitle
\section{Introduction}
Thermoelectrics are currently an object of immense interest and intense research activity owing to the possible enhancement 
of the energy conversion efficiency via nano-structuring \cite{Millie_1,Millie_2,Chen_1} and novel materials design \cite{Heremans_DOS}. 
A higher energy conversion efficiency of a thermoelectric system is typically quantified by an increase in $zT$, the dimensionless {\it{figure of merit}}. 
The figure of merit $zT$ is defined as:
\begin{equation}
zT=\frac{S^2\sigma T}{\kappa_{el}+\kappa_{ph}}, 
\label{eq:zt}
\end{equation} 
where $S,\sigma$, and $\kappa_{el,(ph)}$ are the linear response transport coefficients, namely the thermopower (Seebeck coefficient), 
the electrical conductivity and the electron (phonon) thermal conductivity, with $T$ being the average operating temperature. The proposed increase in $zT$ 
is envisioned via novel approaches towards engineering the electronic \cite{Millie_1,Millie_2,Chen_1,Heremans_DOS} or phononic transport \cite{Tritt_2,Tritt_1,Chen_2} properties. 
Among various low-dimensional nanoscale systems, zero-dimensional systems such as quantum dots have been of special interest because they may exhibit 
an infinitely high value of $zT$ in the absence of phonon thermal conductivity \cite{Mahan_Sofo}. 
\\ \indent The energy conversion efficiency $\eta$ of a thermoelectric system is usually defined as $\eta=\frac{P}{J_{Q}^{in}}$, 
with $P$ being the extracted power and $J_{Q}^{in}$ being the input heat current. Consider, for example, a set-up with a central system sandwiched between two 
reservoirs held at a fixed temperature and electrochemical potential. Under the assumption of small electrochemical potential, $\Delta \mu$, and small temperature, $\Delta T$, differences 
between the reservoirs, the electric current $J$, and the heat current $J_{Q}$ may be written as \cite{Callen,Johnson}:
\begin{eqnarray}
J = L_{11}\Delta \mu + L_{12} \Delta T, \nonumber\\
J_{Q}= L_{21} \Delta \mu + L_{22} \Delta T,
\label{def_lin_resp}
\end{eqnarray}
where $L_{ij}$ represent the Onsager coefficients. The Onsager coefficients $L_{ij}$, are in turn related to the linear response parameters, 
namely $\sigma$, $S$, and $\kappa_{el}$, that appear in the aforementioned definition of $zT$. The efficiency $\eta$, 
when maximized with respect to $J$, yields its maximum $\eta_{max}$ to be an increasing function of $zT$ \cite{Ioffe} given by:
 \begin{equation}
 \eta_{max}=\eta_C \frac{\sqrt{1+zT}-1}{\sqrt{1+zT}+\frac{T_C}{T_H}},
 \label{eq:zt_eff}
 \end{equation}
where the thermoelectric material operates between two contacts maintained at temperatures $T_H$ and $T_C$, with $\eta_C=1-\frac{T_C}{T_H}$ being the Carnot efficiency. 
It is therefore convenient to employ $zT$ as a performance metric to facilitate the design of maximally-efficient thermoelectrics.
\\ \indent However, as noted above, the use of $zT$ as the performance metric in lieu of the actual efficiency relies on the assumption of linear response. 
While a high figure of merit $zT$ is often a necessary component for a good thermoelectric, it does not sufficiently underpin the working conditions that are involved. 
For example, an analysis of the figure of merit $zT$ of the single orbital quantum dot system \cite{Mahan_Sofo}, under the condition of vanishing coupling to the contacts, 
simply points to its infinite value and the resulting efficiency maximum as the Carnot value. It was pointed out only recently \cite{Linke_1,Linke_2,Linke_3,PhysRevB.82.045412,Espo_1,Sanchez} 
that this efficiency maximum only occurs under {\it{open circuit}} conditions, implying an operating condition with a vanishing current and hence a vanishing power output. 
The Carnot efficiency is reached only due to the possibility of this reversible operation \cite{Linke_1,Linke_3} under open circuit conditions. 
\\ \indent The open circuit condition, although associated with a vanishing current, is an operating point which has 
both an electrochemical potential gradient and a temperature gradient. The voltage $V_S$ associated with this electrochemical potential difference $\pm q V_S$, 
with $q$ being the electric charge, is known as the Seebeck voltage. This voltage cancels the current set up by the applied temperature gradient. It is hence pertinent to analyze 
thermoelectric operation by using a nanocaloritronic set up, wherein the central system is subject to a bias drop, not necessarily equal to the Seebeck voltage, 
and a temperature gradient.
\\ \indent The central system considered in our nanocaloritronic analysis is a single orbital interacting quantum dot. Steady state non-equilibrium currents through the central system, 
rather than linear response parameters, are used to evaluate the power and hence the efficiency at each operating point. Each operating point is defined by the applied bias 
and the applied temperature gradient. The primary goal of our transport calculations is to identify the operating conditions that 
point to a specific operating efficiency in relation to the operating power. Some recent works \cite{Espo_1,Espo_2,Espo_3,Linke_2}, for example, have specifically analyzed 
the operation of a single orbital non-interacting quantum dot thermoelectric set up at maximum power. A recent investigation that includes Coulomb interactions \cite{PhysRevB.82.045412} 
has noted the importance of non-linear effects, and has focused on the role of a phonon bath on the thermoelectric operation. Another recent work \cite{mukherjee_moore} 
has focused on the effect of Coulomb interaction on the figure of merit $zT$. The main focus of this paper therefore is a comprehensive performance analysis 
of a quantum dot thermoelectric set up.  The following fundamental aspects of the thermoelectric operation of a quantum dot set up are then discussed in detail: 
a) With a finite coupling to the contacts, a thermoelectric set up always operates irreversibly under open-circuit conditions, with a zero efficiency.
b) Operation at a peak efficiency close to the Carnot value is possible under a finite power operation. In the non-interacting single orbital case, 
the peak efficiency approaches the Carnot value as the coupling to the contacts becomes smaller. In the interacting case, this trend depends non-trivially on the interaction parameter $U$. 
c) The evaluated trends of the maximum efficiency derived from the non-equilibrium currents deviate considerably from the conventional {\it{figure of merit}} $zT$ based result. 
We point out in detail the discrepancies between our non-equilibrium analysis, and the linear response analysis that is usually based on the figure of merit $zT$.
Given the current experimental possibility of thermoelectrics across zero-dimensional systems \cite{Reddy_1,Reddy_2}, and the recent theoretical activity 
exploring non-linear thermoelectric effects \cite{Linke_1,Linke_2,PhysRevB.82.045412,Espo_1,Espo_2,Espo_3} across them, our paper elucidates 
the importance of Coulomb interaction on their thermoelectric performance. 
\\ \indent This paper is organized as follows. Section II describes the necessary formulation: first the definition of the electric and energy currents through the quantum system, 
then the formalism used to evaluate these currents and hence the power output and efficiency across it. The quantum transport system 
under consideration is a single level Anderson-impurity type quantum dot that is weakly coupled to the contacts in the sequential tunneling limit. 
The formulation for currents follows from the density matrix master equation approach under this sequential tunneling approximation \cite{Konig_1,Brouw_1,Timm,Grifoni_1}. 
Section III begins by describing the thermoelectric operation of a quantum dot set up in the absence of interactions $(U=0)$. 
Following that, the fundamental results due to the introduction of Coulomb interactions (finite $U$) are discussed in detail. 
The section concludes with an analysis of the maximum power operation. It is shown that with Coulomb interactions 
the maximum power operation is relatively unaffected in comparison with the non-interacting case discussed in other works \cite{Espo_1,Espo_2,Espo_3}. 
Section IV summarizes the results of this work.
\section{Theoretical Formulation}
A prototype nanocaloritronic configuration of a quantum thermoelectric set up is shown in Fig.~\ref{fig:def}(a). 
In this set up, the central quantum system is coupled to two reservoirs/contacts $\alpha$, which are labeled $\alpha=H$, (hot) and $\alpha=C$ (cold), 
each of which is characterized by a temperature $T_{H(C)}$ and an electrochemical potential $\mu_{H(C)}$. 
This set up closely relates to that of a heat engine commonly studied in classical thermodynamics. 
Under this set up, the central system plays the role of the heat engine with the reservoirs acting as heat sources or sinks. 
In the case of a thermoelectric set up, however, we also need to invoke the additional concept of {\it{particle exchange}} \cite{Linke_1}, 
due to the fact that the reservoirs are characterized by both temperature and electrochemical potential, thus facilitating both energy and particle transport. 
An applied voltage bias $qV_{app}=\mu_C-\mu_H$, an applied temperature gradient $\Delta T=T_H-T_C$, or both, 
trigger particle flux and hence a flow of both an electric and an energy current results. In order to describe quantum thermoelectric transport 
across the system of interest, electric, and energy currents must be clearly defined.  
\subsection{Electric and energy currents}
\begin{figure}[tbp]
	\centering
		\includegraphics[width=3.3in,height=3.0in]{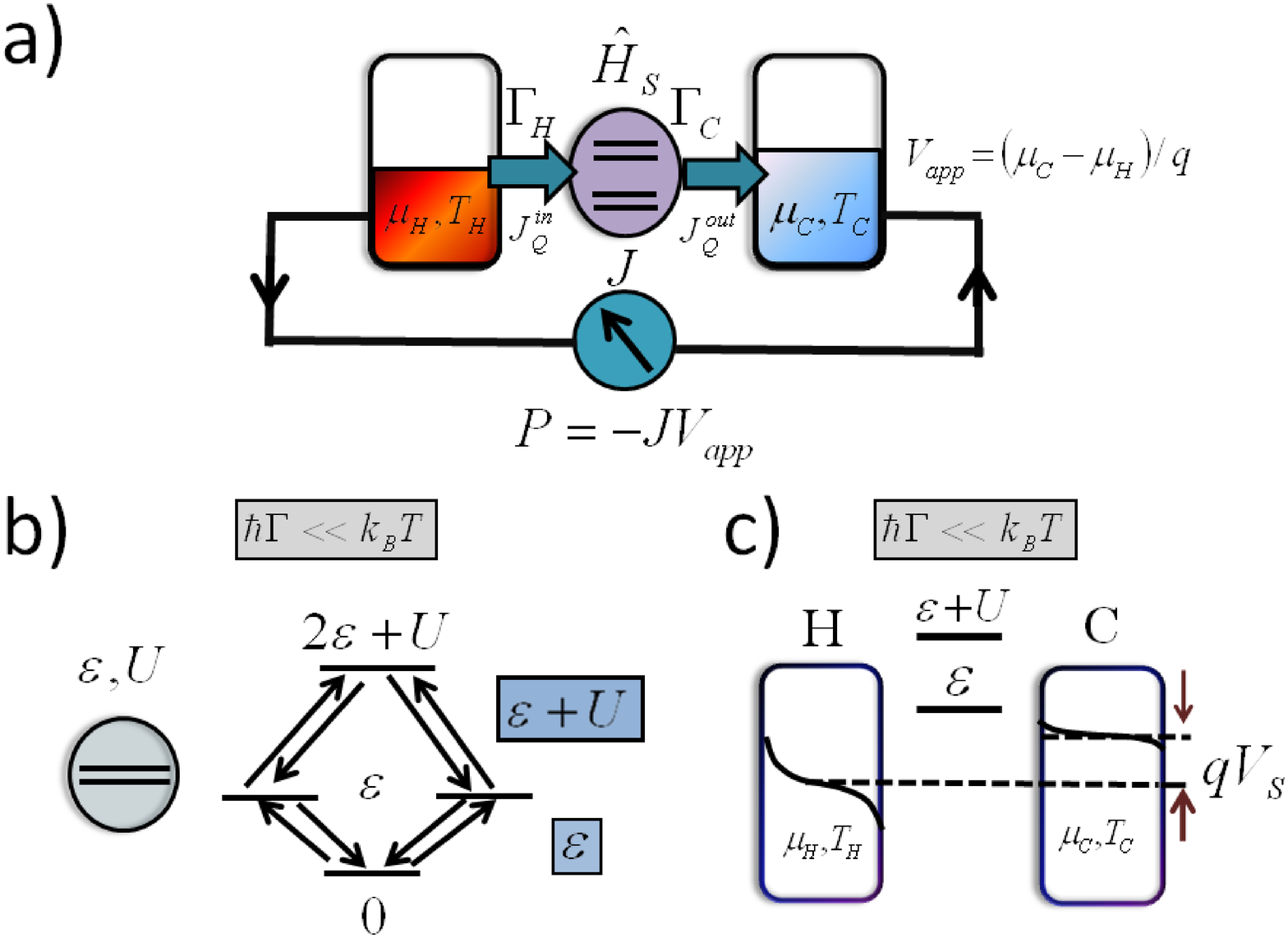}
		\caption{(Color online) Nanocaloritronics of a quantum thermoelectric transport set up. 
a) A typical thermoelectric set up comprises of the central quantum system described by the Hamiltonian $\hat{H}_S$
sandwiched between two reservoirs labeled hot(cold) $\alpha=H(C)$. 
When this central system is subject to an electrochemical potential gradient and a temperature gradient, the resulting current $J$ drives an electrical power $P=J V_{app}$ 
via the electrical leads. Equal contact couplings $\Gamma_H=\Gamma_C$ are assumed throughout. 
b) A single orbital quantum dot is parameterized by its single particle energy level $\epsilon$ and 
the Coulomb interaction parameter $U$. Transport is represented as transitions between states of the many-particle spectrum with electron numbers differing by $\pm 1$. Transport channels 
then comprise the energy difference $\epsilon, \epsilon+U$ between those states with electron numbers differing by $\pm 1$. 
c) Schematic depicting the thermoelectric effect under open circuit conditions: The built-in or Seebeck voltage $V_S$ enforces zero current in the circuit. 
Useful power can be extracted when the applied voltage $V_{app} \in [0,V_{S}]$, where the condition $V_{app}=V_{S}$ enforces open circuit operation. 
The thermoelectric efficiency defined in the operating region $V_{app}\in [0,V_{S}]$ is strongly affected by the energy difference $\epsilon-\mu_{\alpha}$, 
the applied temperature gradient $\Delta T=T_H-T_C$, and the magnitude $U$ of the Coulomb interaction. } 
\label{fig:def}
\end{figure}

We begin with the fundamental thermodynamic equation relating the internal energy $E$ of a non-magnetic system with extensive variables such as entropy $S$, volume $V$, 
and  particle number $N$ given by \cite{Morandi}: 
\begin{equation}
E(S,V,N)=T S -p V + \mu N,
\label{def_heat}
\end{equation}
where the intensive variables are the temperature $T$, pressure $p$, and the electrochemical potential $\mu$. 
The above definition relates to the thermodynamic state of the system. The case of thermoelectric transport involves electron transfer processes during 
which the thermodynamic state of the system changes. When such processes are involved, one measures the change in the internal energy with respect to the change in the extensive parameters. 
Specific to our case, thermoelectric transport occurs at constant volume $(\Delta V=0)$. Electron transfer processes occur between either reservoir $(\alpha=H,C)$, each maintained at a
fixed temperature and a fixed electrochemical potential, and the system as shown in Fig.\ref{fig:def}(a). 
One may then write an equation for the infinitesimal change in the internal energy due to an infinitesimal charge transfer between either reservoir $(\alpha=H,C)$ and the system as:
\begin{equation}
 dE_{\alpha}=T_{\alpha}dS_{\alpha}+\mu_{\alpha} dN_{\alpha}.
\label{def_heat_chg}
\end{equation}
Using the above definition, we can take total time derivatives to define a current associated with the corresponding flux of the extensive variables given by: 
\begin{equation}
J^{\alpha}_E=T_{\alpha} J^{\alpha}_S + \mu_{\alpha} J^{\alpha}_N.
\label{def_heat_curr}
\end{equation}
The quantity $T_{\alpha}J^{\alpha}_S$ is usually termed as the {\it{heat current}} involved in the isothermal electron transfer between either reservoir and the system. 
This quantity, in general, is the contribution to the energy current that keeps track of entropy flow as given by $J_S$. 
Although the term {\it{heat current}} is widely used in literature, it may not by itself be accurate because heat is not a state function and the definition of 
differentials of such quantities may not be obvious. However, in the case of thermoelectric transport that is considered in this work, the quantity $J_Q$ may be termed heat current, following the arguments put forward in early 
works \cite{Callen,Johnson}. The central assumption is that the reservoirs are maintained in equilibrium and hence the flow of charge and heat to and from the reservoirs happens reversibly. 
All irreversible processes are expected to occur in the interfacial region between the reservoirs and the quantum dot. 
In such a reversible process, the expression $\Delta Q_{\alpha} = T_{\alpha} \Delta S_{\alpha}$ holds true, for the reservoir. 
Hence, the equation may now hence be recast in terms of the quantity $J^{\alpha}_{Q}=T_{\alpha}J^{\alpha}_S$ as:
\begin{equation}
J^{\alpha}_E=J^{\alpha}_{Q} + \mu_{\alpha} J^{\alpha}_N,
\label{def_heat_curr_2}
\end{equation}
where, the quantity $J^{\alpha}_Q$ is the heat current in the reservoir $\alpha$.
In the quantum mechanical case, in order to define currents, we define the time-dependent average current due to an operator $\hat{O}$ 
that is associated with one of the extensive variables as follows:
\begin{eqnarray}
J_O (t) &=& \left < \frac{d \hat{O}}{dt} \right >, \nonumber\\
\frac{d \hat{O}}{dt} &=& -\frac{i}{\hbar}[\hat{H},\hat{O}]+\frac{\partial \hat{O}}{\partial t},
\label{def_curr_quant}
\end{eqnarray}
where $[\hat{H},\hat{O}]$ represents the commutator of the overall Hamiltonian $\hat{H}$ with the operator $\hat{O}$. For a quantum mechanical set up, 
based on the schematic of Fig.~\ref{fig:def}(a), the description of currents thus begins by describing the overall Hamiltonian $\hat{H}$ 
which is usually written as $\hat{H}=\hat{H}_S + \hat{H}_{R} + \hat{H}_{T}$, where $\hat{H}_S, \hat{H}_R$ and $\hat{H}_{T}$ represent the system, 
reservoir and reservoir-system coupling Hamiltonians respectively. In this paper, the system comprises of the single orbital Anderson impurity-type 
quantum dot subject to Coulomb interaction described by the following one-site Hubbard Hamiltonian:
\begin{equation}
\hat{H}_S=\sum_{\sigma} \epsilon_{\sigma} \hat{n}_{\sigma} + U \hat{n}_{\uparrow} \hat{n}_{\downarrow},
\label{Ham_def}
\end{equation}
where $\epsilon_{\sigma}$ represents the orbital energy, $\hat{n}_{\sigma} = \hat{d}_{\sigma}^{\dagger} \hat{d}_{\sigma}$ is the occupation number operator of an electron with 
spin $\sigma=\uparrow$, or $\sigma=\downarrow$, and $U$ is the Coulomb interaction between electrons of opposite spins occupying the same orbital. The exact-diagonalization 
of the system Hamiltonian then results in four Fock-space energy levels labeled by their total energies 
$0,\epsilon_{\uparrow},\epsilon_{\downarrow}$ and $\epsilon_{\uparrow}+\epsilon_{\downarrow} +U$. 
In this paper, we consider only a spin-degenerate level such that $\epsilon=\epsilon_{\uparrow}=\epsilon_{\downarrow}$. 
Electronic transport generally involves the addition and removal of electrons. 
In the limit of weak contact coupling $(\hbar \Gamma \ll k_BT)$, transport may be viewed as of transitions between the 
Fock-space levels that differ by an electron number of $\pm 1$ as shown in Fig.~\ref{fig:def}(c). 
The reservoir/contact Hamiltonian is given by 
$\hat{H}_{R} = \sum_{\alpha=H,C}\sum_{k \sigma} \epsilon_{\alpha k \sigma} \hat{n}_{\alpha k \sigma}=\sum_{\alpha k \sigma} \hat{h}_{\alpha k \sigma}$, 
where $\alpha$ labels the hot/cold reservoir ($H$ or $C$ in our case) and the summation is taken over the single particle states labeled $\{k \sigma \}$. 
The tunneling Hamiltonian represents the system-contact coupling usually written as 
$\hat{H}_{T}=\sum_{\alpha k \sigma} \left ( t_{\alpha k \sigma} \hat{c}^{\dagger}_{\alpha k \sigma} \hat{d}_{\sigma} + {{t}^{\ast}_{\alpha k \sigma}} \hat{d}^{\dagger}_{\sigma} 
\hat{c}_{\alpha k \sigma} \right ) = \sum_{\alpha k \sigma} \hat{h}_{T\alpha k \sigma}$, 
where $(\hat{c}^{\dagger},\hat{c})$ and $(\hat{d}^{\dagger},\hat{d})$ denote the creation/annihilation operators of the reservoir and system states respectively.
\\ \indent Pertinent to our problem, one can use Eq.(\ref{def_curr_quant}) to evaluate, for example the steady-state electric, and 
energy currents through the system. The steady state current is then derived in the limit when $t \rightarrow \infty$. Also, in our case 
the operator does not explicitly depend on time implying that $\frac{\partial \hat{O}}{\partial t}=0$. We can then write the particle current due 
to either contact $\alpha=H/C$ by summing contributions over its one-electron states labeled 
$\{k \sigma \}$ as $J^{\alpha}_N= \left <\sum_{k \sigma} \hat{J}^{\alpha}_{k \sigma} \right >= \left < \sum_{k \sigma} \frac{d \hat{n}_{\alpha k \sigma}}{dt} \right >$. 
The expression for the electric current due to either contact given by $J^{\alpha}=-qJ_N^{\alpha}$ then becomes:
\begin{equation}
J^{\alpha} = -q \left < \sum_{k \sigma} \sum_{k' \sigma'} - \frac{i}{\hbar}[\hat{h}_{T\alpha k' \sigma'},\hat{n}_{\alpha k \sigma}] \right >,
\label{def_curr_part}
\end{equation}
where $q$ is the electronic charge. Likewise, the energy current due to the contact $\alpha$ is written as:
\begin{eqnarray}
J_{E}^{\alpha} &=& \left < \sum_{k \sigma} \frac{d {\hat{h}_{\alpha k \sigma}}}{dt} \right > \nonumber\\
 &=& \left < \sum_{k \sigma} \sum_{k' \sigma'} - \frac{i}{\hbar} [\hat{h}_{T\alpha k' \sigma'},{\hat{h}_{\alpha k \sigma}}] \right >\nonumber\\
&=& \left < \sum_{k \sigma} \epsilon_{\alpha k \sigma} \frac{d {\hat{n}_{\alpha k \sigma}}}{dt} \right >.
\label{def_heat_curr_quant}
\end{eqnarray}
The quantity $J_{Q}^{\alpha}$ due to the contact $\alpha$ then becomes:
\begin{equation}
J_{Q}^{\alpha} = T_{\alpha}J_{S}^{\alpha}=\left < \sum_{k \sigma} (\epsilon_{\alpha k \sigma}-\mu_{\alpha}) \frac{d {\hat{n}_{\alpha k \sigma}}}{dt} \right >.
\label{def_heat_curr_fin}
\end{equation}
The above expression is the commonly employed relationship that connects the so called heat currents with entropy and particle currents \cite{PhysRevB.82.045412}. Calculation of the total 
time derivative of the number operator $\hat{n}_{\alpha k \sigma}$ involves the evaluation of its commutator with the 
tunneling Hamiltonian $\hat{h}_{T \alpha k \sigma}$, as described in Eq.(\ref{def_curr_part}). Following Eq.(\ref{def_curr_quant}), the expectation value of 
an operator is evaluated by tracing over the composite system-reservoir density matrix i.e., $\left < \hat{O} \right > = \mathrm{Tr}\{\hat{\rho}(t) \hat{O}(t) \}$. 
The time evolution of $\hat{\rho}(t)$ is given by the Liouville equation. The reduced density matrix $\hat{\rho}_{red}(t)$ of the system 
may be obtained by performing a trace exclusively over the reservoir space. 
An expansion of the Liouville equation to the second order in the tunneling Hamiltonian in the limit of 
weak contact coupling $(\hbar \Gamma \ll k_BT)$, leads to the density matrix master equation for the reduced density matrix of the system \cite{Konig_1,Brouw_1,Timm}.  
In second order, coherences vanish for the considered single orbital model \cite{Grifoni_1}, and one is 
left with a scalar rate equation \cite{Beenakker,Deshmukh,Muralidharan} in
terms of the occupation probabilities $P^N_i=\langle N, i \mid \hat{\rho}_{red}(t) \mid N, i \rangle $ of each $N$ electron
Fock state $|N,i\rangle$ with total energy $E^N_i$. The index $i$ here labels the states within the $N$ electron subspace. 
This Pauli-master equation then involves transition rates $R_{(N,i)\rightarrow(N\pm
1,j)}$ between states $|N, i \rangle$, and $|N \pm 1, j \rangle$ differing by a single electron, leading to a set
of independent equations defined by the size of the Fock space:
\begin{equation}
\frac{dP^N_i}{dt} = \sum_j [R_{(N\pm 1,j)\rightarrow(N,i)}P^{N\pm 1}_j-R_{(N,i)\rightarrow(N\pm 1,j)}P^N_i],
\label{ebeenakker}
\end{equation}
along with the normalization equation $\sum_{i,N}P^N_i = 1$. Notice that, in the stationary limit considered here, where $t \rightarrow \infty$, the Markov approximation 
implicit in Eq.(\ref{ebeenakker}) becomes exact \cite{Timm,Grifoni_1}. At energies close to the Fermi level, metallic contacts 
can be described using a constant density of states, parameterized using 
the bare-electron tunneling rates $\gamma_{\alpha}=\sum_{k \sigma} \frac{2 \pi}{\hbar} |t_{\alpha k \sigma, s}|^2 \delta(E-\epsilon_{k \sigma})$, with $(\alpha=H/C)$. 
We define the rate constants as:
\begin{eqnarray}
\Gamma_{\alpha ij}^{Nr} &=& \gamma_{\alpha}|\langle N,i|\hat{d}^\dagger_{\sigma}|N-1,j\rangle|^2,  \nonumber\\
\Gamma_{\alpha ij}^{Na} &=& \gamma_{\alpha}|\langle N,i|\hat{d}_{\sigma}|N+1,j\rangle|^2.
\end{eqnarray}
The transition rates for the removal $(|N,i \rangle \rightarrow |N - 1,j \rangle)$, and addition $(|N,i \rangle \rightarrow |N+1,j \rangle)$ transitions are then given by
\begin{eqnarray}
R_{(N,i)\rightarrow(N-1,j)} &=&
\sum_{\alpha=H,C}\Gamma_{\alpha ij}^{Nr}\left[1-f \left( \frac{\epsilon^{Nr}_{ij}-\mu_{\alpha}}{k_BT_{\alpha}} \right )\right],
\nonumber\\
R_{(N,i)\rightarrow(N+1,j)} &=&
\sum_{\alpha=H,C}\Gamma_{\alpha ij}^{Na}f \left ( \frac{\epsilon^{Na}_{ij}-\mu_{\alpha}}{k_BT_{\alpha}}\right ).
\label{eq:total_rate}
\end{eqnarray}
The contact electrochemical potentials and temperatures are respectively labeled as $\mu_{\alpha}$ and $T_{\alpha}$, and $f$ is 
the corresponding Fermi-Dirac distribution function with single particle removal and addition
transport channels given by 
\begin{eqnarray}
\epsilon^{Nr}_{ij} = E^N_i - E^{N -1}_j,  \nonumber \\ 
\epsilon^{Na}_{ij} =E^{N+1}_j - E^N_i.
\label{eq:def_tc}
\end{eqnarray}
Finally, the steady-state solution to Eq.(\ref{ebeenakker}), set by $\frac{dP^N_i}{dt}=0$, is used to obtain
the terminal current associated with contact $\alpha$:
\begin{eqnarray}
J^{\alpha} = -q\sum_{N=1}^{N_{tot}}\sum_{ij}[R^{\alpha}_{(N- 1, j)\rightarrow(N,i)}P^{N-1}_j \nonumber\\
   -R^{\alpha}_{(N,i)\rightarrow(N - 1,j)}P^N_i ],
\label{eq:curr_exp}
\end{eqnarray}
where $N_{tot}$ is the total number of electrons in the system. In our case, for example, $N_{tot}=2$. 
Likewise, the quantity $J_{Q}^{\alpha}$, associated with either contact can be similarly defined using Eq.(\ref{def_heat_curr_fin}) as:
\begin{eqnarray}
J^{\alpha}_{Q} =\sum_{N=1}^{N_{tot}}\sum_{ij}[(\epsilon^{\left (N-1 \right )a}_{ji}-\mu_{\alpha})R^{\alpha}_{(N - 1, j)\rightarrow(N,i)}P^{N - 1}_j \nonumber \\ 
-(\epsilon^{Nr}_{ij}-\mu_{\alpha})R^{\alpha}_{(N,i)\rightarrow(N-1,j)}P^N_i ].
\label{eq:curr_exp_2}
\end{eqnarray}
Here, the sum over reservoir indices $(k\sigma)$ in Eq.(\ref{def_heat_curr_fin}), has been replaced by indices $(i,j)$ corresponding to the system states because of elastic electron transfer 
between the reservoir and the system, described by the energetics $\epsilon_{k\sigma}=\epsilon_{ji}^{(N-1)a}$, for the additive transition, and  $\epsilon_{k\sigma}=\epsilon_{ij}^{Nr}$, 
for the removal transition. Notice from Eq.(\ref{eq:total_rate}), that the total rates $R_{(N,i)\rightarrow(N \pm 1,j)}$, and $R_{(N \pm 1,j)\rightarrow(N,i)}$ 
appearing in Eq.(\ref{ebeenakker}), are the sum of individual rates associated with either contact in Eq.(\ref{eq:curr_exp}) and (\ref{eq:curr_exp_2}). 
\subsection{Power and efficiency}
In a classical heat engine, the efficiency of a thermodynamic cycle is defined as $\eta=\frac{W}{Q_{in}}$, which is simply the ratio between 
the work extracted and the heat supplied. However, while working with the nanocaloritronic configuration shown in Fig. \ref{fig:def}(a), 
it is important to evaluate the efficiency under a finite power operation because conversion of entropy currents to electric currents is desired. 
In our case, in which the operation at a finite power is desired, the efficiency is given in terms of the rates of flow of various quantities: 
\begin{equation}
\eta=\frac{P}{J_{Q}^{in}},
\label{eq:def_eff}
\end{equation}
where, the instantaneous power or just the power is defined as $P=(J_{Q}^{in}-J_{Q}^{out})$. 
Following Eq.(\ref{def_heat_curr_fin}), and assuming no intra-system or 
endo-dynamic energy changes due to inelastic processes, the net electrical power between the hot and cold reservoirs can be written as:
\begin{equation}
P= \left ( J_{Q}^{H}+J_{Q}^{C} \right )= -\frac{1}{q}(\mu_C-\mu_H) J = -V_{app}  J,
\label{eq_power}
\end{equation}
where $J=J^H=-J^C$ refers to the electric current whose magnitude is conserved in steady state. It must be noted that the 
above expression has both the Joule (irreversible) and the thermoelectric (reversible) components \cite{Datta_Lake,Snyder_Ursell}. 
For example, specific to the linear response case, one obtains $P=L_{11} \left ( (\Delta \mu)^2 + L_{12} \Delta T \Delta \mu \right )/q$, by employing Eq.(\ref{def_lin_resp}). 
This combines linear and quadratic terms in the applied voltage bias $qV_{app}=(\mu_C-\mu_H) = \Delta \mu$, the linear term being the thermoelectric part, 
and the quadratic term being the Joule part.
\\ \indent The power generated, and hence the efficiency, is generally evaluated at an operating point. Each operating point is specified by 
the applied bias $V_{app}$ and the temperature gradient $\Delta T=T_H-T_C$. For the upcoming analysis, we work with 
the convention that the temperature gradient is applied at the contact labeled $H$, and the voltage bias $V_{app}$ is applied at the contact labeled $C$. 
In all our calculations, we assume that half of the applied voltage drops across the quantum dot as a 
result of equal capacitive coupling to the two contacts.
\section{Thermoelectric operation of a quantum dot}
In the realm of molecular electronics or quantum dot transport, it is common to start with a microscopic understanding of 
transport processes across a single spin degenerate orbital subject to Coulomb interactions. Often this leads to a 
qualitative physical picture of various experimental observations and the additional complexity of multiple levels may append mainly to the quantitative aspect. 
Based on the formulation discussed in the previous section, we first elucidate the thermoelectric operation of the quantum dot set up without Coulomb interactions. 
Following that, we discuss the important results of this work that arise due to the inclusion of Coulomb interactions. 
\subsection{Power and efficiency of a non-interacting quantum dot thermoelectric set up} 
First we discuss the results that follow from the sequential tunneling model. 
This model implies a delta line shape for the quantum dot density of states and transmission function.  
In this limit, the analytical result for the currents are given by:
\begin{eqnarray}
J&=& \frac{-2q \gamma_H \gamma_C}{\gamma_H+ \gamma_C} \left( f_H(\epsilon)-f_C(\epsilon) \right), \nonumber \\
J_{Q}^{\alpha}&=& \frac{2\gamma_H \gamma_C}{\gamma_H+ \gamma_C} (\epsilon-\mu_{\alpha})\left( f_H(\epsilon)-f_C(\epsilon) \right), 
\label{eq:curr_one_level}
\end{eqnarray}
with $\gamma_{H,C}$ being the contact coupling energies associated with contacts $H,C$. Here, $f_{\alpha}(\epsilon)=f \left ( \frac{\epsilon-\mu_{\alpha}}{k_BT_{\alpha}} \right )$ 
refers to the Fermi-Dirac distribution of either contact. The factor of $2$ appears due to spin-degeneracy in the non-interacting case. 
\\ \indent Based on the schematic in Fig.~\ref{fig:def}(c), the basic thermoelectric operation can be described as follows. An electric current is set up by 
the applied temperature gradient. Under open circuit conditions, the Seebeck voltage $V_{S}$ is set up in order to oppose this current. 
This built-in voltage can be used to drive power across an electrical system, say a resistor. Alternatively, while working in a circuit configuration an externally 
applied voltage bias $V_{app}$ may be used as a variable electric current source. The condition that enforces zero electric current 
is then equivalent to an operating condition with an applied bias $V_{app}=V_{S}$. It is now easy to see from Eq.(\ref{eq:curr_one_level}) that a zero electric current may be 
enforced by $f_H(\epsilon)=f_C(\epsilon)$ or 
\begin{equation}
\frac{\epsilon-\mu_H}{k_BT_H}=\frac{\epsilon-\mu_C}{k_BT_C}.
\label{eq:cond_open}
\end{equation}
In general, the quantity $J_{Q}^{\alpha}$ under the above condition need not also be identically zero. In the present case, however, $J_{Q}^{\alpha}$ is also zero, and is easily noted from Eq.(\ref{eq:curr_one_level}). 
This point has an important implication with respect to the operating efficiency.
\\ \indent It can be shown by using the definition of efficiency Eq.(\ref{eq:def_eff}) and Eq.(\ref{eq:curr_one_level}), that the efficiency is given by:
\begin{equation} 
\eta=\frac{(\mu_C-\mu_H)}{(\epsilon-\mu_H)}. 
\label{eq:eff_single_dot}
\end{equation}
Notice that the expression for the efficiency is 
independent of the current $J$. Under open circuit conditions, it can then be deduced that the maximum efficiency $\eta_{max}=\eta_{C}$. Thus, {\it{a non-interacting quantum dot, under the limit of 
vanishing coupling to the contacts, operates reversibly and achieves the Carnot efficiency under open circuit conditions}}. A brief discussion of the thermodynamic aspects of this reversible 
operation is carried out in Appendix I. Finally, making another connection with \cite{Mahan_Sofo}, 
the zero value of the quantity $J_{Q}^{\alpha}$ results in a zero electron thermal conductivity, i.e., $\kappa_{el}=\frac{J_{Q}}{\Delta T}=0$. This results in an infinitely 
high value of $zT$ in the absence of phonon contribution, and following Eq.(\ref{eq:zt_eff}), leads to the maximum efficiency equaling that of the Carnot efficiency.
\begin{figure}
	\centering
		\includegraphics[width=3.16in,height=3.6in]{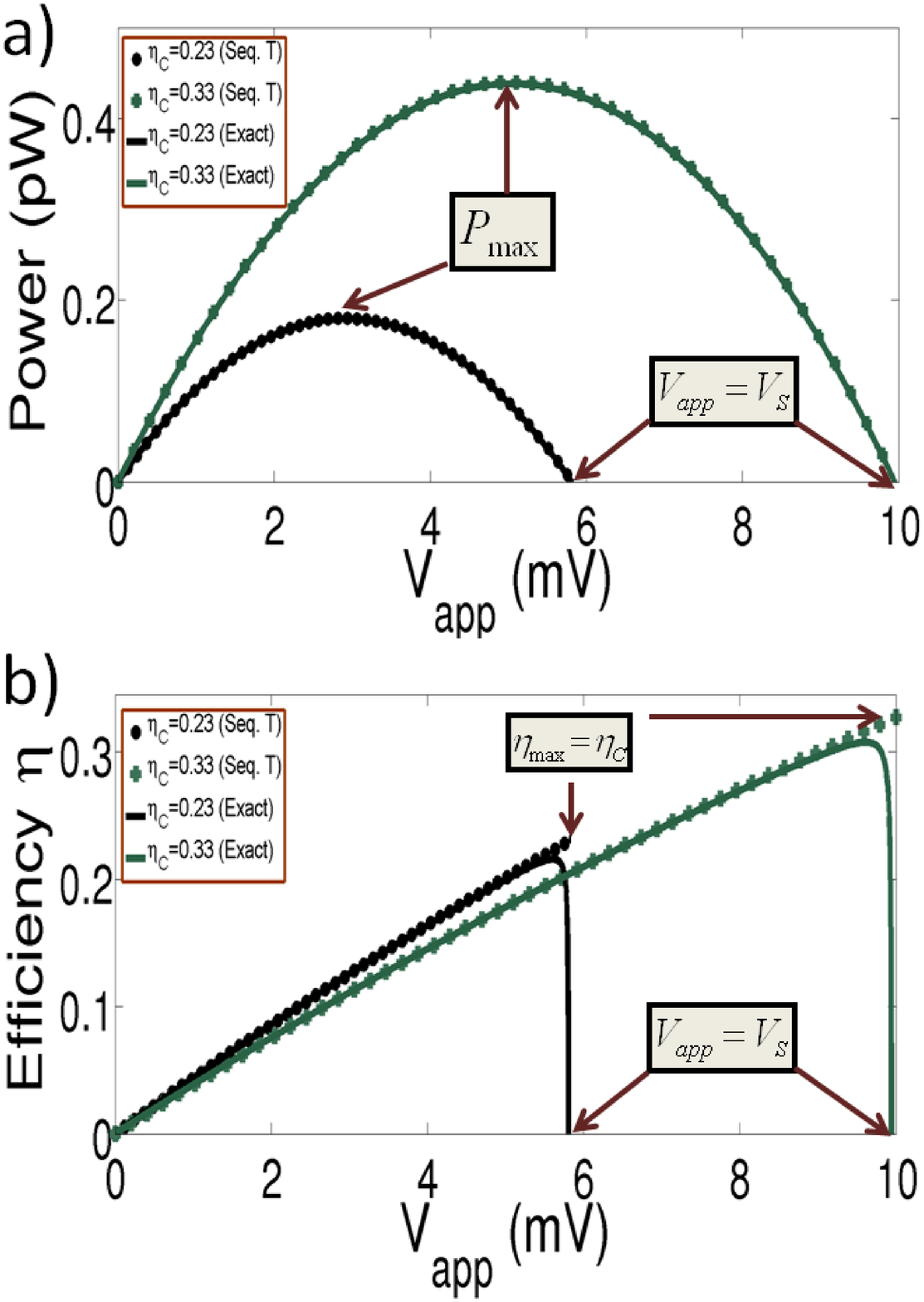}
		\caption{(Color online) Power and efficiency in the non-interacting $(U=0)$ limit for Carnot efficiency $\eta_C=0.23$ (black solid) and $\eta_C=0.33$ (green circles). The temperature at the 
cold contact is set to $T_{C}=100K$, and the equilibrium energy level placement is set to $\epsilon-\mu_H=2k_BT_H$ at $V_{app}=0$. The couplings to the reservoirs are 
taken as $\hbar \gamma_H= \hbar \gamma_C=0.01 meV \approx 10^{-3}k_BT$. a) Plot of extracted power as a function of the applied bias $V_{app}$. The span of the operating region $V_{app} \in [0,V_{S}]$ 
broadens with increase in the applied temperature gradient $\Delta T =T_H-T_C$. Results from the sequential tunneling approximation (dotted) and the exact calculation (bold) are identical. 
b) Corresponding plots of efficiency in the operating region. Under the sequential tunneling approximation (dotted), the efficiency maximizes at the Carnot efficiency $\eta_C$ when the applied bias equals the built-in voltage $(V_{app}=V_{S})$. 
This corresponds to the reversible thermoelectric configuration \cite{Mahan_Sofo,Linke_1,Linke_2} (see text). In the exact calculation, however (bold), 
the efficiency drops to zero under open circuit conditions. The efficiency at maximum power lies in an intermediate operating point corresponding to the maximum power $P_{max}$ shown in (a).} 
\label{fig:def_op_reg}
\end{figure}
\\ \indent {\it{Effect of line width:}} The rather surprising result of achieving a finite efficiency under zero power operation is indeed an artifact of the sequential tunneling 
approximation which implies the idealized delta form for the transmission function. Going beyond the sequential tunneling approximation, the delta function broadens, 
and hence impacts the conclusions drawn above. In the specific case of the non-interacting limit, it is possible to exactly 
evaluate the currents using for example, the transmission formalism \cite{Linke_2} as:
\begin{eqnarray}
 J&=& \frac{-2q \gamma_H \gamma_C}{\gamma_H+ \gamma_C} \int_{-\infty}^{\infty}dE D(E)\left( f_H(E)-f_C(E) \right), \nonumber \\
J_{Q}^{\alpha}&=& \frac{2\gamma_H \gamma_C}{\gamma_H+ \gamma_C} \int_{-\infty}^{\infty} dE D(E) (E-\mu_{\alpha})\left( f_H(E)-f_C(E) \right),\nonumber \\
\label{eq:landauer} 
\end{eqnarray}
where the broadened density of states $D(E)$ is given by:
\begin{equation}
 D(E)= \frac{1}{2 \pi} \frac{\gamma_H + \gamma_C}{\left ( \left (E-\epsilon \right )^2 + \left ( \left (\gamma_H + \gamma_C \right )/2 \right )^2 \right )}.
\label{eq:DOS}
\end{equation}
The efficiency, as seen in Fig.~\ref{fig:def_op_reg}(b), given by the ratio of $P$ and $J_{Q}^H$, drops to zero under open circuit conditions. 
This is because unlike in the previous case, it can be noted from Eq.(\ref{eq:landauer}), that $J^H_Q$ need not also vanish when $J=0$. This also implies that the Carnot efficiency can never be reached in the real situation. 
In our simulations, we have used a $\hbar \gamma_H = \hbar \gamma_C= 0.01 meV = 10^{-3} k_BT$, such that the condition for weak coupling to the contacts $\hbar \Gamma \ll k_BT$, is satisfied. 
One can hence note from Fig.~\ref{fig:def_op_reg}(a) that there is almost no difference in the variation of power between sequential tunneling approximation and the exact calculation. 
The sequential tunneling limit thus provides a very good approximation for the evaluation of currents in the limit of weak coupling to the contacts, 
but fails to describe the correct trend for the efficiency in this limiting case. 
\\ \indent The efficiency under open circuit conditions is identically zero, because the quantity $J^H_Q$ is finite when the current $J$ is zero.  
Using Eq.(\ref{eq:landauer}), and $J_{Q}^{\alpha}=T_{\alpha} J_S^{\alpha}$, we note that:
\begin{equation}
 J^C_S-J^H_S = \int_{-\infty}^{\infty} dE \jmath(E) \left ( \frac{(E-\mu_{C})}{T_C} - \frac{(E-\mu_{H})}{T_H} \right ),
\label{eq:entropy_curr}
\end{equation}
where, $\jmath(E)=\frac{2\gamma_H \gamma_C}{\gamma_H+ \gamma_C} D(E)\left( f_H(E)-f_C(E) \right)$. Physically, the above result implies that 
although the flow of electrons from the hot to the cold contact under open circuit conditions is balanced by the reverse flow, the net flow of entropy is not. 
Entropy can then be produced while maintaining a zero net particle flux. This entropy production thus results in a finite entropy current under open circuit conditions. 
Therefore, it implies that unlike the special case of vanishing coupling to the contacts that is discussed in Appendix I, 
spontaneous electron exchange between the reservoirs is inherently irreversible. The entropy generated is dissipated as heat deep in the reservoirs.
\\ \indent {\it{Operating region:}} Thus far, we have considered only one operating condition, namely that of the open circuit operation in which $V_{app}=V_{S}$. 
In order to fully characterize the thermoelectric set up, an understanding of its operation at an arbitrary applied bias $V_{app}$ must be considered. 
According to our convention, and following the definition of efficiency in Eq.(\ref{eq:def_eff}), useful work may be extracted only in the region of positive power 
$(P \ge 0)$. Therefore, the domain in which $P \geq 0$, that is represented by the applied bias $0 \leq V_{app} \leq V_{S}$, as shown in the schematic in Fig.~\ref{fig:def}(a), 
defines the operating region. The extracted power in this operating region $V_{app} \in [0,V_{S}]$ is shown in Fig.~\ref{fig:def_op_reg}(a), for two different values of the applied temperature gradient, and hence 
of the Carnot efficiency. Notice that the extracted power is identically zero $(P=0)$ under two operating conditions: short circuit condition- when $V_{app}=0$, and open circuit condition- 
when $V_{app}=V_{S}$. The first one corresponds to a zero bias and the second one corresponds to the condition with a zero electric current. The operating region 
also becomes larger as the applied temperature gradient $\Delta T$, and consequently the associated Carnot efficiency $\eta_C=\frac{\Delta T}{T_H}$ is increased. 
This is because an increase in $\Delta T$ increases the amount of current flowing through the level, as a result of which a higher applied voltage $V_{app}=V_{S}$ 
is needed to counter it. The variation of power in the operating region is quasi-quadratic and has a maximum (marked $P_{max}$) in the operating region.
 \\ \indent The efficiency in the sequential tunneling case increases monotonically and quasi-linearly \cite{Espo_2,Sanchez} from $0$ to $\eta_C$ in the operating region. 
In the exact calculation, however, the efficiency reaches a maximum that is close to the Carnot value and then drops to zero at the open circuit operation point $V_{app}=V_S$. 
The abruptness of this behavior depends on how large the coupling to the contacts is. This deviation of the maximum efficiency obtained via the exact calculation from the ideal Carnot value 
obtained via the sequential tunneling approximation will become more pronounced as the contact coupling is increased.
\subsection{Power and efficiency of an interacting quantum dot thermoelectric set up}
With the same initial configuration as in the previous case, we now study the effect of varying $U$. 
Referring to the state transition diagram in Fig.~\ref{fig:def}(b), the transport spectrum now consists of 
the addition and removal levels $\{\epsilon \}=\{\epsilon^{1a}_{00},\epsilon^{1r}_{00}\}$, where $\epsilon^{1r}_{00} = E^1_0 - E^{0}_0=\epsilon$, 
and $\epsilon^{1a}_{00} = E^{2}_0 - E^1_0=\epsilon +U$. The expressions for the steady state currents \cite{PhysRevB.82.045412,Deshmukh,mukherjee_moore} 
through the hot contact (say) $\alpha=H$, based on Eqs.(\ref{def_heat_curr_fin}) and (\ref{eq:curr_exp}), are given by:
\begin{eqnarray}
J^{H} &=& -q\left ( R^{H}_{0 \rightarrow 1} P^0 -R^{H}_{1 \rightarrow 0} P^1+ R^{H}_{1 \rightarrow 2} P^1 -R^{H}_{2 \rightarrow 1} P^2 \right ), \nonumber\\
J_{Q}^{H} &=& \left (\epsilon-\mu_{H} \right ) \left ( R^{H}_{0 \rightarrow 1} P^0 -R^{H}_{1 \rightarrow 0} P^1 \right ) \nonumber\\
 \qquad &+& \left (\epsilon+U-\mu_{H} \right ) \left ( R^{H}_{1 \rightarrow 2} P^1 -R^{H}_{2 \rightarrow 1} P^2 \right ),
\label{eq:curr_CB}
\end{eqnarray}
where $P^{N}$'s are the occupation probabilities of the many body state with $0$, $1$ or $2$ electrons. 
We have dropped the index $i$ within each $N$ electron subspace, because only ground states exist within the framework of our spin-degenerate single orbital system.
The solution for the set of master equations for this system based on Eq.(\ref{ebeenakker}) is straightforward, 
and yields the following expressions for the occupation probabilities:
\begin{eqnarray}
P^0=\frac{1}{\Omega}R_{1 \rightarrow 0} R_{2 \rightarrow 1}, \nonumber \\
P^1=\frac{1}{\Omega}R_{0 \rightarrow 1} R_{2 \rightarrow 1}, \nonumber \\
P^2= \frac{1}{\Omega} R_{0 \rightarrow 1} R_{1 \rightarrow 2},
\label{eq:prob}
\end{eqnarray}
with $\Omega$ being the normalization factor that ensures the sum of probabilities to be equal to unity. Here the total rate 
$R_{i \rightarrow j}=\sum_{\alpha} R^{\alpha}_{i \rightarrow j}$, given by the sum of the rates due to each contact $\alpha=H,C$. 
To be specific, the addition rates due to contact $\alpha=H$ in Eq.(\ref{eq:curr_CB}) are given by $R^{H}_{0 \rightarrow 1}=\gamma_{H}f(\epsilon-\mu_{H})$, 
and $R^{H}_{1 \rightarrow 2}=\gamma_{H}f(\epsilon+U-\mu_{H})$, and the removal rates are given by $R^{H}_{1 \rightarrow 0}=\gamma_{H}(1-f(\epsilon-\mu_{H}))$, 
and $R^{H}_{2 \rightarrow 1}=\gamma_{H}(1-f(\epsilon+U-\mu_{H}))$. 
  \begin{figure}
	\centering
		\includegraphics[width=3.25in,height=3.8in]{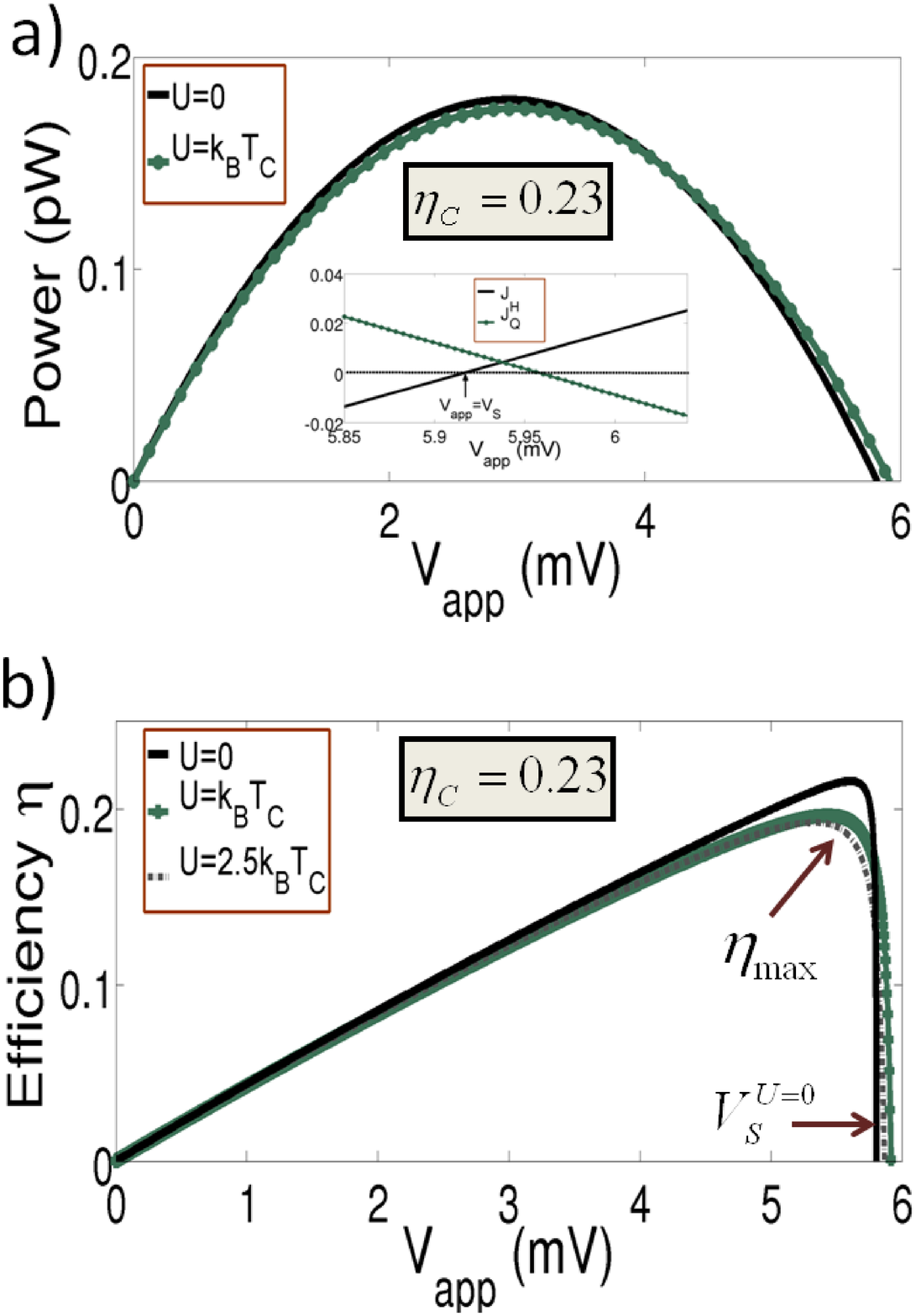}
		\caption{(Color online) Power and efficiency at finite $U$ for $\eta_C=0.23$. a) Power extracted in the operating region. 
The span of the operating region in the case of $U=k_BT_C$ (green circles) can be different from that of the non interacting case (black solid). 
In general, the quantity $J_{Q}^{H}$ (see inset) is not identically zero when the electric current vanishes. 
b) Variation of the efficiency in the operating region for different values of $U$: (i) $U=0$ (black solid), (ii) $U=k_BT_C$ (green solid) and (iii) $U=2.5k_BT_C$ (gray dashed). 
Note that, with finite $U$ such that $U > \hbar \Gamma$, the efficiency is identically zero when the electric current vanishes under open circuit conditions $(V_{app}=V_{S})$. 
The efficiency also reaches a maximum $\eta_{max}$ at finite power operation.}
	\label{fig:eta_var_U}
\end{figure}
\\ \indent We now plot the power (Eq.(\ref{eq_power})), and efficiency (Eq.(\ref{eq:def_eff})) in the operating region in Fig.~\ref{fig:eta_var_U}. 
In comparison with the non-interacting case, the domain of the operating region in the finite $U$ case (green circles) is slightly different. 
This is because a finite $U$ introduces a transport channel at $\epsilon+U$ in addition to the already existing one at $\epsilon$ as shown in Fig.~\ref{fig:def}(c). 
For very small values of the interaction parameter $U$, specifically when $U \approx \hbar \Gamma$, 
higher order tunneling processes may become relevant. Such processes may only be captured by a perturbative expansion beyond 
the second order in the tunneling Hamiltonian \cite{Grifoni_1}.
\\ \indent We now plot the variation of the efficiency along the operating region $V_{app}$ in Fig.~\ref{fig:eta_var_U}(b) 
for different values of $U$. The trend of the variation of the efficiency with finite $U$ is similar to what was noted in the non-interacting case. 
The efficiency reaches a maximum $\eta_{max}$ before becoming zero. However, we also note from  Fig.~\ref{fig:eta_var_U}(b) that the abruptness of this variation 
is less stark in comparison with the non-interacting case. In other words the maximum efficiency $\eta_{max}$ occurs well within the domain of finite power. 
The introduction of interactions therefore also results in {\it{maximum efficiency within a finite power operation}}.
This observed trend of the efficiency with applied voltage as noted in Fig.~\ref{fig:eta_var_U}(b) may be qualitatively understood by analyzing 
the variation of currents with the applied voltage $V_{app}$. Based on Eq.(\ref{eq:curr_CB}), one may recast an expression for the currents as:
\begin{eqnarray}
J &=& -q( J_1(\epsilon) + J_2(\epsilon+U) ), \nonumber\\
J_{Q}^{H} &=& (\epsilon-\mu_{H})J_1(\epsilon) + (\epsilon +U -\mu_{H})J_2(\epsilon+U), \nonumber \\
\label{eq:curr_CB_2}
\end{eqnarray}
where $J_1(\epsilon)$ and $J_2(\epsilon+U)$ denote the contribution 
to the electric currents due to the transport channels at $\epsilon$ and $\epsilon+U$ and are given by:
\begin{eqnarray}
J_1(\epsilon) &=& \frac{\gamma_H \gamma_C R_{2 \rightarrow 1}}{\Omega}(f_H(\epsilon)-f_C(\epsilon)), \nonumber\\
J_2(\epsilon+U) &=& \frac{\gamma_H \gamma_C R_{0 \rightarrow 1}}{\Omega}(f_H(\epsilon+U)-f_C(\epsilon+U)). \nonumber \\
\label{eq:curr_CB_3}
\end{eqnarray}
Likewise,
\begin{equation}
 J_{Q}^{C} = (\epsilon-\mu_{C})J_1(\epsilon) + (\epsilon +U -\mu_{C})J_2(\epsilon+U).
\label{eq:curr_heat_cold}
\end{equation}
When $V_{app}$ is large enough to allow double occupancy in the quantum dot, the second transport channel $\epsilon+U$ begins to conduct. The electrical current then 
redistributes between the two transport channels. From Eq.(\ref{eq:curr_CB_2}), we note that the magnitude of $J_{Q}^{H}$ becomes more prominent as 
the contribution $J_2(\epsilon+U)$ increases. This causes the $J_{Q}^{H}$ to approach the zero value less rapidly with increasing bias than how the electric current would, 
thereby resulting in an overall decrease in the ratio $\frac{J}{J_{Q}^H}$ between them. This causes the efficiency $\eta=\frac{-JV_{app}}{J_{Q}^H}=\frac{P}{J_{Q}^{H}}$ to decrease 
with increasing $V_{app}$ once the maximum $\eta_{max}$ is reached. The applied bias at which this happens depends on $U$, and the above effect of the 
second transport channel will vary as $U$ is increased. 
\\ \indent An important consequence of the introduction of this extra transport channel at $\epsilon+U$, is that both the currents defined in Eq.(\ref{eq:curr_CB}) do not vanish 
at the same operating point. Here, as shown in the inset of Fig.~\ref{fig:eta_var_U}(a), $J_{Q}^{H}$ is finite even when the electric current $J$ 
vanishes when $(V_{app}=V_{S})$. The open circuit condition from Eq.(\ref{eq:curr_CB_2}), can be deduced as $J_1(\epsilon)=-J_2(\epsilon+U)$. It then follows from Eqs.(\ref{eq:curr_CB_3}) and (\ref{eq:curr_heat_cold}), 
that $J^{\alpha}_Q=T_{\alpha} J_S^{\alpha}=UJ_2(\epsilon+U)$, and hence
\begin{equation}
 J^C_S-J^H_S =  \left ( \frac{U}{T_C}- \frac{U}{T_H} \right ) J_2(\epsilon+U).
\label{eq:entr_chg_tot}
\end{equation}
It is thus noted that similar to the non-interacting case, under open circuit conditions, a net entropy generation occurs, thus making spontaneous 
electron transfer processes irreversible.
\begin{figure}
	\centering
		\includegraphics[width=3.25in,height=3.8in]{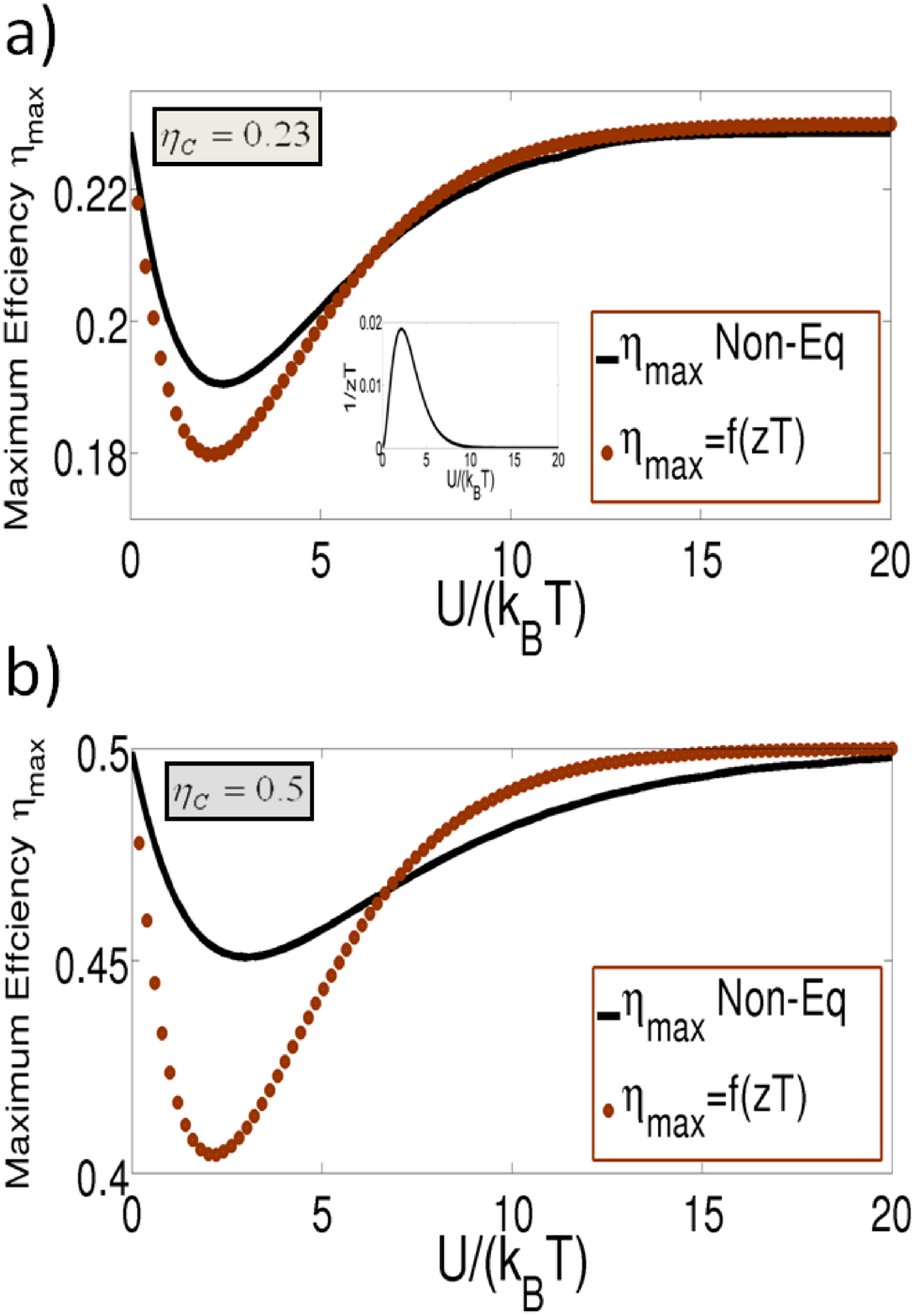}
		\caption{(Color online) Variation of the maximum efficiency with Coulomb interaction $U$. The maximum efficiency is equal to the Carnot efficiency for 
$U = 0$ and asymptotically approaches it when $U \gg k_BT$. It reaches a minimum around $U \approx 2.7k_BT$. This variation is shown for a) $\eta_C=0.23$ and b) $\eta_C=0.5$. 
Also shown in each case is the comparison between the non-equilibrium calculation (bold) and that based on the figure of merit $zT$ (brown dotted). Note that the difference between them 
becomes more prominent for larger values of $\eta_C$ or larger temperature gradients $\Delta T$, thereby making the transport non-linear and hence the concept of $zT$ 
less useful. The inset in (a) shows the variation of $1/zT$ with $U$ for the chosen level configuration $\epsilon-\mu_H=2k_BT_H$ at $V_{app}=0$.} 
	\label{fig:eta_max_U_comp}
\end{figure}
\\ \indent In order to further probe as to how the interaction $U$ influences the achievable maximum efficiency, we plot the variation of $\eta_{max}$ (shown bold) with 
$U$ in Fig.~\ref{fig:eta_max_U_comp}. We notice that with increasing $U$, the maximum efficiency reaches its global minimum around $U \approx 2.7 k_BT$, and asymptotically 
approaches Carnot efficiencies at very large values of $U$. As $U$ is increased beyond $U \approx 2.7 k_BT$, the second transport channel $\epsilon+U$ becomes less accessible, 
and transport resembles the previous case with only one transport channel $\epsilon$. Thus, the important implication here is that the variation of maximum efficiency 
with the introduction of interactions is non-trivial and non-monotonic. 
\\ \indent Our results are based on the evaluation of non-equilibrium currents and hence go beyond linear response. It is hence desirable to compare our results 
directly with the conventional $zT$ based evaluation which is valid only in the linear response limit. Using Eq.(\ref{eq:zt_eff}), the linear response 
maximum efficiencies calculated from $zT$ are also plotted in  Figs.~\ref{fig:eta_max_U_comp}(a) and (b),(brown circles).  
It must be noted from Fig.~\ref{fig:eta_max_U_comp}(a) and (b) that the non-equilibrium calculation deviates from the $zT$ based calculation \cite{mukherjee_moore} (brown circles), 
and that this discrepancy is more pronounced for larger values of the Carnot efficiency. Also, comparing Fig.~\ref{fig:eta_max_U_comp}(a) and (b), it is seen that the deviation of the 
non-equilibrium efficiency from the Carnot efficiency with increasing $U$ is less pronounced for larger values of the Carnot efficiency. 
\begin{figure}
	\centering
		\includegraphics[width=3.25in,height=3.85in]{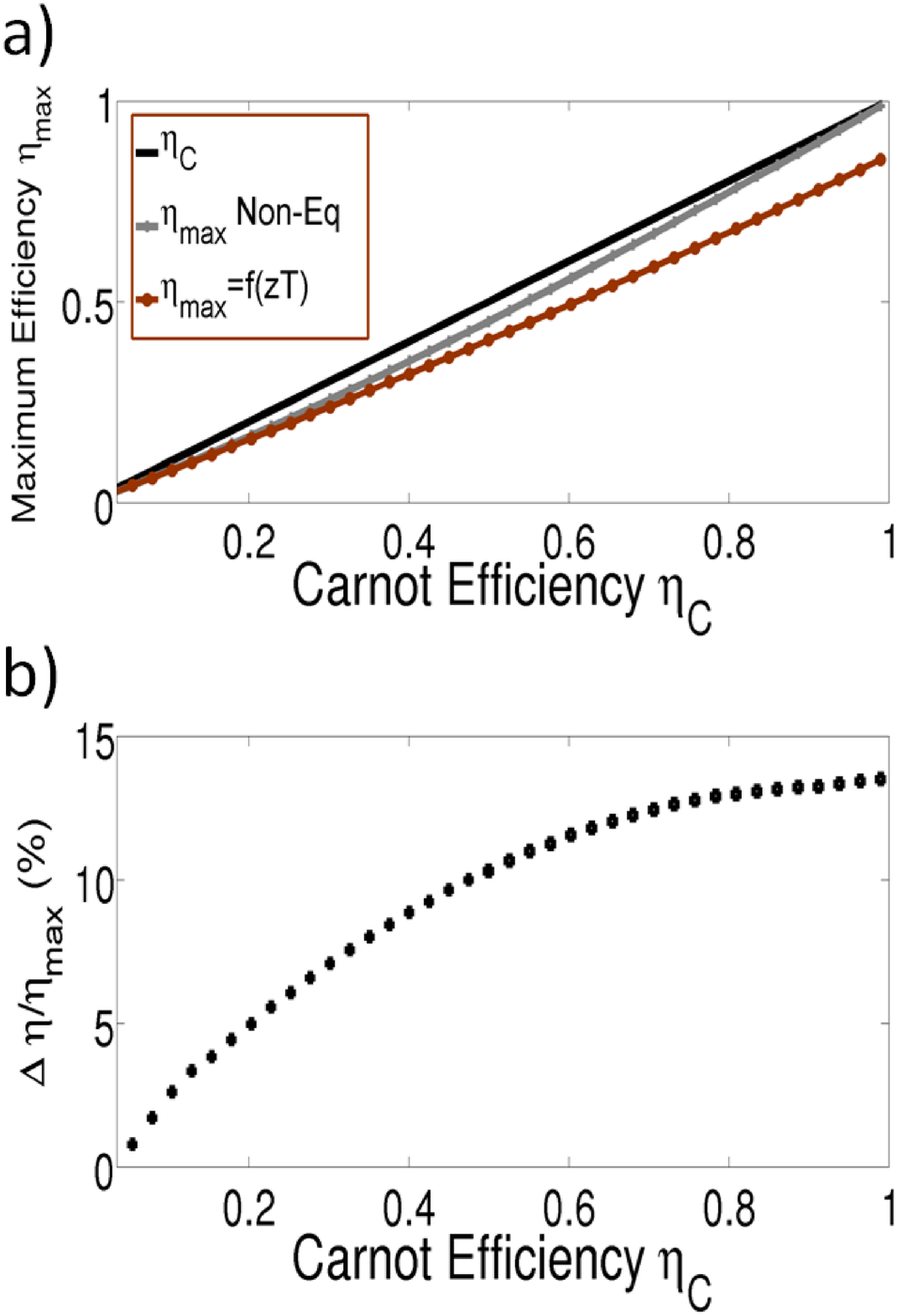}
		\caption{(Color online) Variation of maximum efficiency with respect to $\eta_C$. a) The maximum efficiency (gray dotted) approaches 
the Carnot efficiency and deviates more from the figure of merit $zT$ based calculation (brown dotted) as the Carnot efficiency increases. 
b) Plot of the percentage deviation of maximum efficiency between the non-equilibrium evaluation and the $zT$ based evaluation. The maximum efficiencies 
at each value of $\eta_C$ here are taken from the respective global minimum $(U \approx 2.7 k_BT)$ in their variation with respect to $U$ in Fig.~\ref{fig:eta_max_U_comp}. } 
	\label{fig:eta_max_U_car}
\end{figure} \\ 
\indent To elucidate better, the discrepancy between the non-equilibrium evaluation and a $zT$ based evaluation, 
we plot in Fig.~\ref{fig:eta_max_U_car}, the variation of the non-equilibrium evaluation (gray squares) and the $zT$ 
based evaluation (brown circles) of the maximum efficiency as a function of $\eta_C$, the Carnot efficiency. 
We note from Fig.~\ref{fig:eta_max_U_car}(a) that the non-equilibrium calculation of $\eta_{max}$ deviates less from the Carnot value 
for both small and large values of $\eta_C$, with the maximum deviation in the intermediate region. On the contrary, the $zT$ based calculation 
deviates from both $\eta_C$ and the non-equilibrium evaluation with increasing Carnot efficiency. From Eq.(\ref{eq:zt_eff}), in the $zT$ based evaluation of the maximum efficiency, 
$\eta_{C}$ is modulated by an increasing function of $zT$ and is not strongly dependent on the operating conditions. Thus as $\eta_C$ is increased, 
thereby increasing the applied temperature gradient, non-equilibrium effects become prominent and transport cannot be adequately captured by the $zT$ based calculation. 
The inset in Fig.~\ref{fig:eta_max_U_comp}(a) shows the variation of $1/zT$ with $U$, illustrating that $zT \rightarrow \infty$ in the 
two opposite limits $U \ll k_BT$ and $U \gg k_BT$. The percentage deviation between the non-equilibrium calculation and the $zT$ based calculation of $\eta_{max}$ 
as a function of $\eta_C$ is plotted in Fig.~\ref{fig:eta_max_U_car}(b).
\subsection{Maximum power operation of an interacting quantum dot thermoelectric set up}
While Carnot efficiency poses the ultimate limit for any heat engine, there may or may not be other fundamental limits involved under finite power operation. 
It has been shown that the maximum power operation of any Carnot engine is limited by the Curzon-Ahlborn efficiency $\eta_{CA}=1-\sqrt{1-\eta_C}$ \cite{Curzon}. 
The study of the maximum power operation of a non-interacting quantum dot set up has been pursued previously \cite{Espo_1,Espo_2,Espo_3}. 
Here, we analyze the maximum power operation of the quantum dot system with the inclusion of Coulomb interaction.
\begin{figure}
	\centering
		\includegraphics[width=3.25in,height=1.95in]{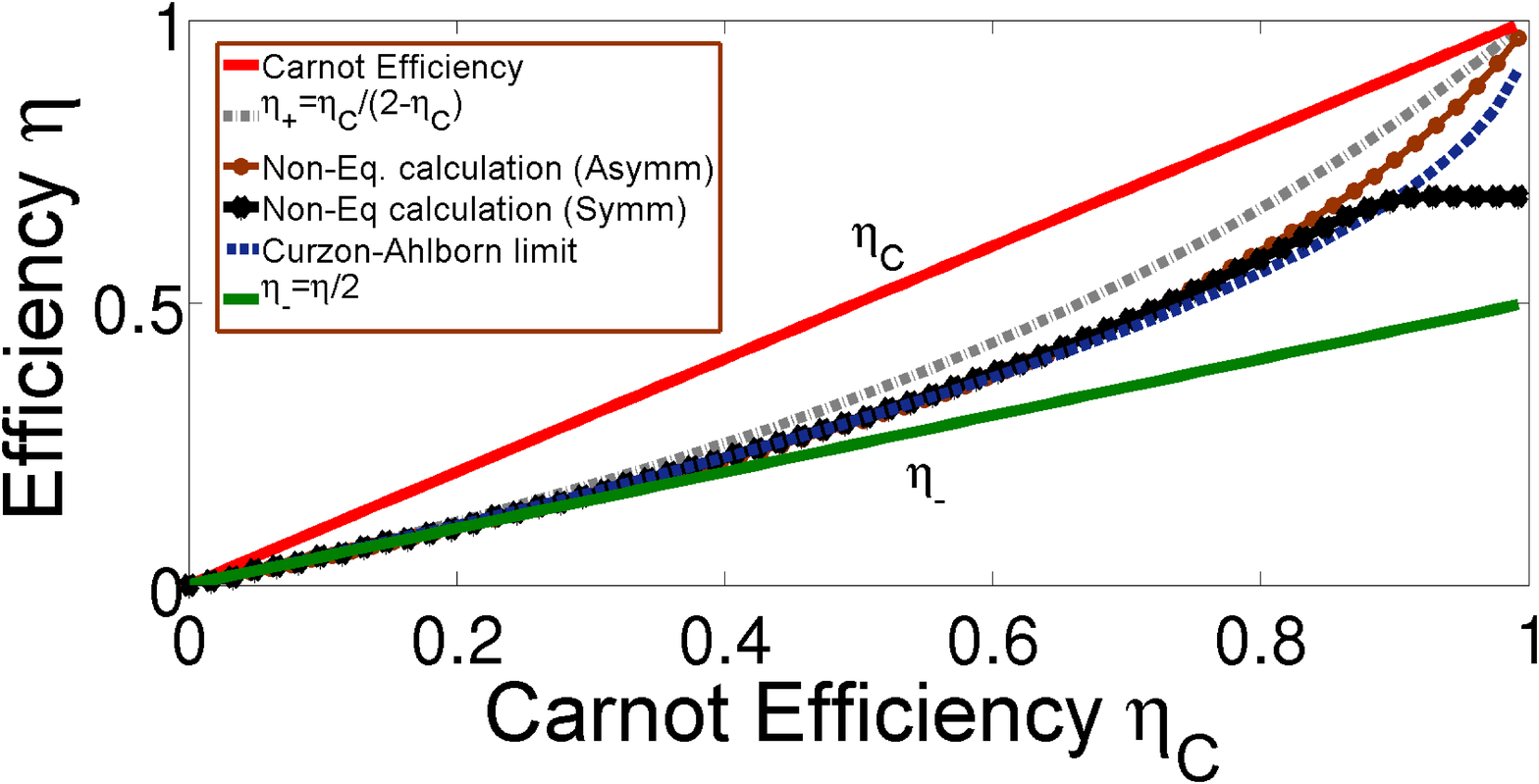}
		\caption{(Color online) Comparison between the efficiency at maximum power and various other limits for the interacting quantum dot set up. 
The efficiency at maximum power is evaluated at $U \approx 2.7 k_BT$, which corresponds to the maximum deviation from $\eta_C$ in Fig.~\ref{fig:eta_max_U_comp}. 
The non-equilibrium evaluation for our set up using both symmetric electrostatic coupling (black diamond), and fully asymmetric electrostatic coupling (brown circles) is shown. 
The non-equilibrium evaluation assuming no electrostatic coupling to the cold contact, resembles the curves discussed in \cite{Espo_1}.
The non-equilibrium evaluation of the efficiency at maximum power, in general, is dependent on the details of the set-up and need not be strictly bound by limits discussed in \cite{Espo_3}. } 
	\label{fig:CA_eff}
\end{figure}
\\ \indent Analyzing maximum power operation implies maximizing the electrical power output $P=-J  V_{app}$. The efficiency at maximum power $\eta_{MP}$ 
is calculated with respect to the operating point that maximizes the power output. We consider how this efficiency at maximum power $\eta_{MP}$ compares 
with various limits discussed in earlier works \cite{Espo_1,Espo_2,Espo_3}. Also, the efficiency at maximum power is evaluated for $U \approx 2.7 k_BT$, which corresponds to 
the maximum deviation of the efficiency from $\eta_C$ in Fig.~\ref{fig:eta_max_U_comp}. We plot in Fig.~\ref{fig:CA_eff} the 
calculation of the quantity $\eta_{MP}$ under two set up conditions: (a) symmetric electrostatic coupling that is used throughout the paper (shown black diamond), and b) fully asymmetric set up in which the 
voltage applied across the cold contact is electrostatically decoupled to the quantum dot (shown brown circles).  
In Fig.~\ref{fig:CA_eff}, we note that for smaller values 
of the temperature difference, and hence smaller values of the Carnot efficiency $\eta_C$, the efficiency at maximum power $\eta_{MP}$ remains close to 
the Curzon Ahlborn limit and is approximately linear. In this limit, the curves follow a linear law. An important observation is that, 
similar to what was inferred in \cite{Espo_1}, the efficiency at maximum power $\eta_{MP}$ need not be bounded by the Curzon-Ahlborn efficiency for 
larger values of the Carnot efficiency $\eta_C$, and may indeed be larger. This questions the regime of applicability of the Curzon-Ahlborn limit, 
which may only be valid for working conditions close to linear response. Consider an expansion for the Curzon-Ahlborn efficiency 
$\eta_{CA}=1-\sqrt{(1-\eta_C)}$ in powers of $\eta_C=\frac{\Delta T}{T_H}$ written as 
\begin{equation}
\eta_{CA}=\frac{\eta_C}{2}+\frac{\eta_C^2}{8} + \ldots, 
\end{equation}
from which it can be noted that for smaller values of $\Delta T$ and hence smaller values of the Carnot efficiency $\eta_C$, the non-equilibrium efficiency 
follows the linear term after which the quadratic term dominates. Notably, the deviation of the non-equilibrium efficiency at maximum power 
with respect to the Curzon-Ahlborn limit in Fig.~\ref{fig:CA_eff} elucidates the fact that this limit need not be a fundamental limit as the Carnot limit is. 
Physically, this implies that under non-equilibrium conditions, the leading term in the power expansion for $\eta_{CA}$ deviates from a non-equilibrium evaluation, 
and importantly is specific to the set up. It has been pointed out in a recent work \cite{Espo_3} that in the limit of low dissipation, the efficiency at maximum power $\eta_{MP}$ for a 
Carnot engine is bounded as $\eta_{-} \leq \eta_{MP} \leq \eta_+$, where $\eta_{-}=\frac{\eta_C}{2}$ with $\eta_{+}=\frac{\eta_C}{(2-\eta_C)}$. Note that our curve of the efficiency at 
maximum power is also not necessarily bound between the above two extrema. 
\\ \indent We thus note that the trend of the efficiency at maximum power shown in Fig.~\ref{fig:CA_eff} is similar to that of the non-interacting case analyzed 
in previous works \cite{Espo_1,Espo_2,Espo_3}, when the quantum dot is electrostatically decoupled with the cold contact. 
The fact that the efficiency at maximum power, under these conditions can approach the Carnot limit at certain 
larger values of $\Delta T$ (and hence $\eta_C$), points out to the possibility of high power operation at high efficiencies. While a large ratio of $\frac{\Delta T}{T_H}$ 
is not feasible at higher operating temperatures, it may be an interesting possibility in low temperature applications.   
\section{Conclusions}
In this paper, we analyzed the performance of an interacting quantum dot thermoelectric set up. This study was based on the evaluation of power and efficiency 
from the non-equilibrium currents in the sequential tunneling limit. The operating region of the thermoelectric set up was identified and a general trend of the efficiency in this 
operating region was identified. We showed that the much discussed aspect 
of reversible operation with Carnot efficiency, under open circuit conditions, in the case of non-interacting single orbital quantum dot system, only occurs in the limit of vanishing coupling to the contacts. 
In a general case, the efficiency reaches a maximum in the operating region before dropping to zero at the open circuit operating point. 
In the non-interacting single orbital case, the efficiency can become very close to the Carnot value, if the coupling to the contacts is sufficiently weak. 
In the interacting case, we showed that this trend depends non-trivially on the interaction parameter $U$.
We also pointed out the clear discrepancy between our non-equilibrium evaluation of the maximum efficiency $\eta_{max}$ and the figure of merit $zT$ based calculation, 
which is only valid in the linear response limit. Comparison of the efficiency at maximum power with the Curzon-Ahlborn limit and other related bounds were also discussed. 
Here, it was shown that the inclusion of Coulomb interactions did not alter the already noted conclusions in the non-interacting case \cite{Espo_1,Espo_2,Espo_3}. However, the trend of 
variation of the efficiency at maximum power is set up dependent. Our current theoretical treatment, however, is in the limit of weak coupling to the contacts, and symmetric contact coupling. 
In the regime of asymmetric and strong contact coupling, we expect novel physics that may be introduced by asymmetric charging \cite{Ferdows,Miller} 
to affect the thermoelectric transport processes. This will be an object of future research and possible extension of the current work.
\\ {\bf{Acknowledgements:}} This work was supported by the Deutsche Forschungsgemeinschaft (DFG) under programmes SFB 631 and SFB 689. 
We thank Christoph Strunk for useful discussions. The author BM acknowledges useful email correspondence with Rafael Sanchez. 
\section{Appendix I: Reversible operation}
The following important aspects of the so called reversible operation in the case of a non-interacting quantum dot set up under vanishing coupling to the contacts must be noted. 
Firstly, the fact that Carnot efficiency is achieved points out to a 
reversible operation in an infinite time thermodynamic cycle. This naturally implies that no power is drawn, although the cycle achieves the highest possible Carnot efficiency 
by performing work for an infinite period of time. Secondly, the term {\it{reversible}} has the following implication with respect to electronic transport. 
Let us consider the entropy generated when an electron is transfered from the hot to the cold reservoir via the energy state $\epsilon$ in the quantum dot. 
This involves the electron transfer (i) from the hot reservoir to the quantum dot, and (ii) from the quantum dot into the cold reservoir. The entropy change in the hot reservoir due 
to process (i) is given by $\Delta S_H=-\frac{\epsilon-\mu_H}{k_BT_H}$, because the hot reservoir has lost an electron of energy $\epsilon$. Similarly, 
the entropy change of the cold reservoir due to process (ii) is given by $\Delta S_C=\frac{\epsilon-\mu_C}{k_BT_C}$, because the cold reservoir has gained an electron of energy $\epsilon$.
Thus, the entropy change per electron for the forward (hot to cold) $(\Delta S_f)$ and the 
reverse (cold to hot) $(\Delta S_r)$ transfer processes between the two reservoirs can be written as:
\begin{eqnarray}
\Delta S_f = \left ( \frac{\epsilon-\mu_C}{T_C}-\frac{\epsilon-\mu_H}{T_H} \right ), \nonumber \\ 
\Delta S_r = \left ( \frac{\epsilon-\mu_H}{T_H}- \frac{\epsilon-\mu_C}{T_C} \right ).
\label{eq:entrop_transf}
\end{eqnarray}
Therefore, as pointed out in \cite{Linke_3}, the open circuit condition in the present case namely 
$\frac{\epsilon-\mu_H}{k_BT_H} =\frac{\epsilon-\mu_C}{k_BT_C}$, ensures zero entropy production in either the forward or the reverse electron transfer process. 
Normally, either forward or reverse transfer processes involve the generation of entropy in the set up as a whole.  
This case, however, implies that given only a single orbital energy $\epsilon$, one can have a unique bias configuration given by the open circuit condition, that can result in a spontaneous 
exchange of electrons reversibly without entropy generation. 
\bibliographystyle{apsrev}	
\bibliography{one_level_TE_bib}		
\end{document}